\documentclass[11pt]{article}
\usepackage{moriond,epsfig}

\begin{document}
\vspace*{4cm}
\title{CLUSTERS OF GALAXIES AS SELF-GRAVITATING SYSTEMS}

\author{ Florence DURRET }
\address{ Institut d'Astrophysique de Paris, 98bis Bd Arago, 75014 Paris,
France}

\author{Ricardo DEMARCO}
\address{ESO, Karl Schwarzshild Str. 2, Garching bei M\"unchen, Germany \\
Institut d'Astrophysique de Paris, 98bis Bd Arago, 75014 Paris,
France}

\author{ Fr\'ed\'eric MAGNARD }
\address{ Institut d'Astrophysique de Paris, 98bis Bd Arago, 75014 Paris,
France}

\author{ Daniel GERBAL }
\address{ Institut d'Astrophysique de Paris, 98bis Bd Arago, 75014 Paris,
France}

\maketitle \abstracts{ Self-gravitating systems such as elliptical
galaxies appear to have a constant specific entropy and obey a scaling
law relating their potential energy to their mass.  These properties
can be interpreted as due to the physical processes involved in the
formation of these structures. Dark matter haloes obtained through
numerical simulations have also been found to obey a scaling law
relating their potential energy to their mass with the same slope as
ellipticals.  Since the X-ray gas in clusters is weakly dissipative,
we have checked the hypothesis that it might verify similar
properties. \\ We have analyzed ROSAT-PSPC images of 24 clusters, and
also found that: 1) the S\'ersic law parameters (intensity, shape and
scale) describing the X-ray gas emission are correlated two by two; 2)
the hot gas in all these clusters roughly has the same specific
entropy; 3) a scaling law linking the cluster potential energy to the
mass of the X-ray gas is observed, with the same slope as for
elliptical galaxies and dark matter haloes. }

\section{Introduction}

\begin{figure}
\begin{center}
\psfig{figure=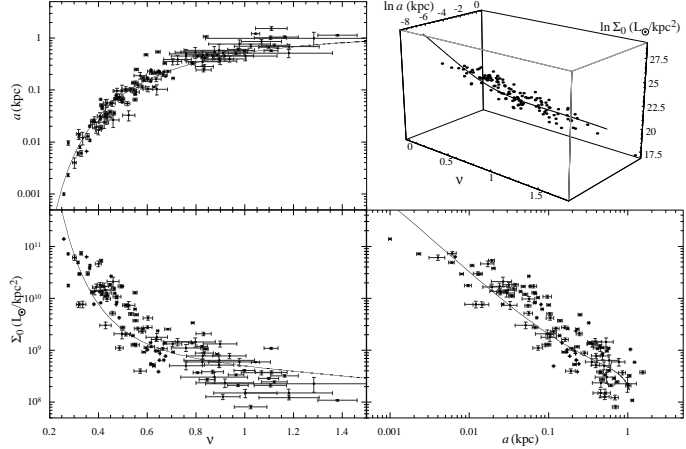,height=6truecm}
\caption{Two by two correlations of the S\'ersic parameters 
$log \Sigma_0 , loga, \nu$ for elliptical galaxies. Distribution of 
elliptical galaxies in the 3D S\'ersic parameter space (top right). 
The lines superimposed in each panel are the correlations predicted 
from the intersection of the Entropic Surface (constant specific entropy)
and Potential energy-mass relation scaling relation.
\label{fig:ficelle}}
\end{center}
\end{figure}

\begin{figure}[h]
\begin{center}
\psfig{figure=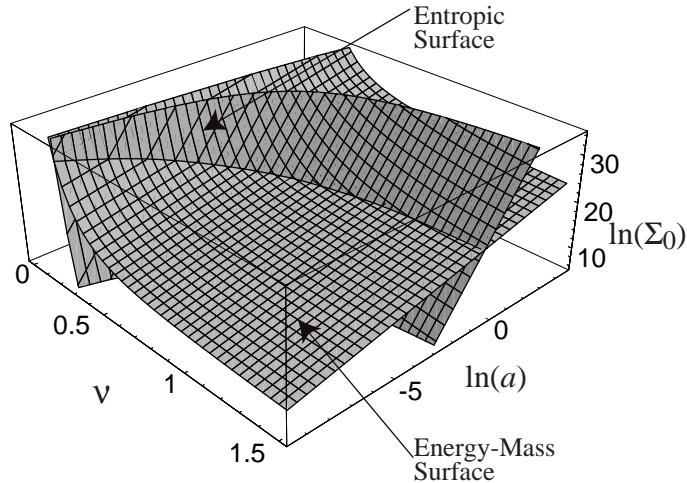,height=6.5truecm}
\caption{3D representation of the specific entropy and potential 
energy-mass two-manifolds for elliptical galaxies, using the coordinates
$[log \Sigma_0 , loga, \nu]$.
\label{fig:persp}}
\end{center}
\end{figure}

The optical surface brightness profiles of elliptical galaxies can
be fit by a S\'ersic law (S\'ersic 1968):

\begin{equation}
\label{sersic_profile}
\Sigma(s)=\Sigma_0\:exp\left[- \left ( \frac{s}{a} \right )^{\nu} \right]
\end{equation}
\noindent
characterized by three parameters: $\Sigma_0$ (intensity), $a$
(scaling) and $\nu$ (shape). We have shown for a sample of 132
ellipticals belonging to three clusters that the S\'ersic parameters
were correlated two by two, and that in the three-dimensional space
defined by these three parameters they were located on a thin line
(Fig. \ref{fig:ficelle}). These properties have been interpreted as
due to the fact that, to a first approximation, all these elliptical
galaxies have the same specific entropy (entropy per unit mass)
(Gerbal et al. 1997, Lima Neto et al. 1999, M\'arquez et al. 2000),
and that a scaling law exists between the potential energy $U$ and the
mass $M$ for these galaxies (M\'arquez et al. 2001): $U\ \propto \
M^{1.72\pm0.03}$. Each of these relations defines a two-manifold in
the [$log \Sigma_0 , loga, \nu$] space.  The thin line on which the
galaxies are distributed in this space is the intersection of these
two two-manifolds (Fig. \ref{fig:persp}).  Such relations are most
probably a consequence of the formation and evolution processes
undergone by these objects, since theory predicts $U\ \propto \
M^{5/3}$ under the hypothesis that energy and mass are conserved
(M\'arquez et al. 2001).

Interestingly, numerical simulations of cold dark matter haloes in two
different mass ranges lead to a similar scaling law between the
potential energy and mass of the haloes. In the mass range $4\ 10^5\
\leq\ M\ \leq\ 4\ 10^8\ M_\odot$ (unvirialized clusters), Jang Condell
\& Hernquist (2001) find a relation consistent with $U\ \propto \
M^{5/3}$, while in the mass range $10^{12}\ \leq\ M\ \leq\ 10^{15}\
M_\odot$ (virialized clusters), Lanzoni (2000) finds $U\ \propto \
M^{1.69\pm0.02}$.

Since the X-ray gas in clusters of galaxies is weakly dissipative, it
is likely to verify similar properties. The work presented here is
meant to check this hypothesis. Our results are still preliminary and
will be fully presented in a forthcoming paper (Demarco et al. in
preparation).

\section{The data and data reduction}

We have selected a sample of 24 clusters with redshifts between $0.01$
and $0.3$ from the ROSAT PSPC archive with a high exposure time (and
therefore signal to noise ratio), a roughly regular structure
(obviously interacting clusters were discarded) and a proper light
curve (no count rate peaks larger than 3 counts/seg in the 4 bands
used). The main properties of the selected clusters are given in Table
\ref{tab:data}.

\begin{table}[h]
\caption{Physical parameters of the 24 clusters of our sample.
\label{tab:data}}
\begin{center}
\begin{tabular}{|rrrrrrrrr|}
\hline
Cluster &  z~~ &$\nu$~~ & $a(kpc)$ & $a_{eq}(kpc)$ & ${n_e}_0(cm^{-3})$ & $p(\nu)$ & $r_{eff}(kpc)$ & $T_0(keV)$ \\
\hline 		            					      	              	                
A85      & 0.0518 &  0.55  &  278   &  252   &      0.00581   &     0.68  &   2939   &   6.20  \\
A478     & 0.0881 &  0.50  &  155   &  138   &      0.01600   &     0.71  &   2572   &   6.90  \\
A644     & 0.0704 &  0.82  &  427   &  371   &      0.00365   &     0.54  &   1216   &   6.59  \\
A1651    & 0.0860 &  0.74  &  411   &  373   &      0.00391   &     0.58  &    1597  &   6.10  \\
A1689    & 0.1810 &  0.58  &  208   &  193   &      0.01407   &     0.67  &   1877   &   9.02  \\
A1795    & 0.0631 &  0.54  &  178   &  157   &      0.01314   &     0.68  &   1939   &   5.88  \\
A2029    & 0.0765 &  0.49  &  164   &  145   &      0.01717   &     0.71  &   2839   &   8.47  \\
A2034    & 0.1510 &  1.00  &  868   &  784   &      0.00179   &     0.45  &   1740   &   7.00  \\
A2052    & 0.0348 &  0.47  &  108   &  96    &      0.01245   &     0.72  &   2359   &   3.10  \\
A2142    & 0.0899 &  0.81  &  601   &  495   &      0.00401   &     0.54  &   1686   &   9.70  \\
A2199    & 0.0299 &  0.60  &  192   &  175   &      0.00733   &     0.65  &   1451   &   4.10  \\
A2219    & 0.2250 &  0.85  &  737   &  667   &      0.00295   &     0.52  &   2017   &   12.40 \\
A2244    & 0.0970 &  0.56  &  237   &  237   &      0.00724   &     0.67  &   2586   &   8.47  \\
A2319    & 0.0559 &  0.80  &  801   &  733   &      0.00197   &     0.55  &   2570   &   9.12  \\
A2382    & 0.0648 &  1.17  &  904   &  797   &      0.00057   &     0.36  &   1398   &   5.00  \\
A2390    & 0.2310 &  0.59  &  347   &  297   &      0.00776   &     0.66  &   2593   &   11.10 \\
A2589    & 0.0416 &  0.72  &  345   &  297   &      0.00253   &     0.59  &   1373   &   3.70  \\
A2597    & 0.0852 &  0.34  &  20.6  &  18    &      0.17712   &     0.80  &   3763   &   4.40  \\
A2670    & 0.0761 &  0.52  &  219   &  213   &      0.00407   &     0.70  &   3127   &   4.45  \\
A2744    & 0.3080 &  1.35  &  949   &  843   &      0.00208   &     0.28  &   1246   &   11.00 \\
A3266    & 0.0594 &  1.18  &  999   &  905   &      0.00126   &     0.36  &   1569   &   8.00  \\
A3667    & 0.0552 &  0.89  &  1143  &  921   &      0.00107   &     0.50  &   2548   &   7.00  \\
A3921    & 0.0960 &  0.81  &  762   &  604   &      0.00150   &     0.54  &   2036   &   4.90  \\
A4059    & 0.0460 &  0.64  &  254   &  222   &      0.00462   &     0.64  &   1479   &   3.97  \\
\hline
\end{tabular}
\end{center}
\end{table}

The data reduction was done using Snowden's now standard software
(Snowden et al., 1994).

\section{The method}

\subsection{Fitting the S\'ersic parameters}

In order to model the X-ray emission of the cluster, we have chosen a
3D density profile $\rho(r)$, which is a generalized form of a
Mellier-Mathez profile (Mellier \& Mathez 1987). It is completely
determined by its three parameters and can be written under the form:

\begin{equation}
\label{mm_profile}
\rho(r)=\rho_0\:{(r/a)}^{-p}\:exp[{-(r/a)}^{\nu}]
\end{equation}
\noindent
where the parameters $p$ and $\nu$ are related by the numerical
aproximation (M\'arquez et al. 2001):

\begin{equation}
\label{p_nu}
p \simeq 1.0 - 0.6097 \nu + 0.05563 {\nu}^2
\end{equation}

This relation gives the best approximation to the S\'ersic law when
equation (\ref{mm_profile}) is projected. The S\'ersic profile defined
by equation (\ref{sersic_profile}) corresponds to a surface mass
density while equation (\ref{mm_profile}) is the volume mass
density. The condition that the mass obtained by integrating equation
(\ref{sersic_profile}) must be equal to the mass obtained by
integrating equation (\ref{mm_profile}), gives us the following
relation between the parameters $\Sigma_0$ and $\rho_0$:

\begin{equation}
\label{rho0_sigma0}
\rho_0=\frac{1}{a}\:\Sigma_0\:\frac{\Gamma(\frac{2}{\nu})}{2\:\Gamma(\frac{3-p}{\nu})}
\end{equation}
where $\Gamma(a)$ is the gamma function defined by $\Gamma(a) = \int^{\infty}_0 x^{a-1} e^{-x} d x$.

In order to determine the correct set of values for the three S\'ersic
parameters for each cluster, we fit the ROSAT images by a
pixel-to-pixel method which: i) creates a three-dimensional model of
the X-ray emission and ii) projects it by integration along the line
of sight, taking into account the energy response and the point spread
function of the detector. The result is a synthetical image which can
be compared with the observation.

To obtain this synthetical image, we have used a code that takes into
account the generalized Mellier-Mathez density profile (equation
(\ref{mm_profile})) and not only the free-free Bremsstrahlung
emission, but also the free-bound and bound-bound X-ray emissions. The
code computes the X-ray emissivity, $\epsilon_{\nu}$, in every point
of the space; this emissivity is then projected by integration along
the line of sight to obtain a surface brightness:

$$\mu(s)=\int^{+\infty}_{z=-\infty}\:\int^{\nu_{max}}_{\nu_{min}}\:\epsilon_{\nu}(s^2+z^2)\:w(\nu)\:d\nu\:dz$$

\noindent
where $s^2=x^2+y^2$, with $x$ and $y$ the perpendicular directions to
the line of sight, and $z$ the coordinate along the line of sight. This
integral is computed taking into account the energy response of the
satellite $w(\nu)$. 

Finally, this projected image is convolved with the ROSAT point spread
function (PSF), which varies as a function of position on the detector
and energy. We have used a FWHM of 2 pixels which corresponds to the
central PSF, because the resulting spreading is more important in the
center where the intensity gradient is stronger.

The cluster redshift and the gas temperature are required as input
parameters.  The redshift for each cluster was taken from the SIMBAD
data base (except for A2199 for which the redshift was obtained from
Wu, Xue \& Fang (1999), and the gas temperature was taken from Wu, Xue
\& Fang (1999), except for A2034 for which the temperature was taken
from Ebeling et al. (1996), and for A2382 for which we have used a
temperature of $5.0\ keV$.  We assumed a standard CDM cosmology with a
Hubble constant $H_0 = 50\:km\:s^{-1}\:Mpc^{-1}$, $\Omega_0 = 1$, and
$\Lambda = 0$.

To obtain an initial guess for the free parameters in equation
(\ref{mm_profile}) we fit a S\'ersic profile (equation
(\ref{sersic_profile})) to the X-ray surface brightness of each
cluster. The ellipticity and semi-major axis position angle of the
X-ray gas were also given as starting values for the code.

Once we have obtained the first synthetic image with the initial guess
for the value of each free parameter, the code compares it to the
actual ROSAT image.  The values of the parameters are then changed and
the comparison process is repeated iteratively, until it finds the 3D
X-ray emission (density dependent) which best fits the surface
brightness profile of the observation when projected. The fitting
process is carried out with the MIGRAD methode in the MINUIT library
of CERN (James 1994).

The best fit parameters for each cluster are: the semi-major axis $a$
(in kpc), the shape parameter $\nu$ (dimensionless) and the central
electronic number density, ${n_e}_0$ (${cm}^{-3}$).  

Since the formula used to calculate the specific entropy is valid only
for spherical symmetry, and in order to take the ellipticity of the
X-ray emission into account, we have also computed a new scale
parameter, $a_{eq}$, defined by $a_{eq}= a\:\sqrt{b/a}$. In this way,
we will calculate the specific entropy of a spherically symmetric
X-ray region with a scale parameter $a_{eq}$.  The best fit parameters
are given in Table~1.

\subsection{Calculation of the various physical quantities}

By integrating equation (\ref{mm_profile}) we find the gas mass as a
function of radius:

\begin{equation}
\label{mass_gas}
M_{gas} (r) = \int^r_0 \rho (u) 4 \pi u^2 d u = \frac{4 \pi \rho_0 a^3}{\nu} \gamma(\frac{3-p}{\nu},(\frac{r}{a})^{\nu})
\end{equation}
where $\gamma(a,z)$ is the incomplete gamma function defined by
$\gamma(a,z) = \int^{z}_{0} x^{a-1} e^{-x} d x$.

In order to estimate the total gas mass, we only have to integrate the
last equation until $r=\infty$ (note that $\rho(r)$ does not
diverge). The result is:

\begin{equation}
\label{mass_gas_tot}
M_{gas}=2 \pi a^2 \Sigma_0 \frac{1}{\nu} \Gamma(\frac{2}{\nu})
\end{equation}
which depends only on the free parameters of the S\'ersic law. The
potential energy of the gas is given by:
\vskip -0.4truecm 
\begin{equation}
\label{u}
U_{pot}\equiv \frac{M^2_{gas}}{R_g}= 4\:\pi^2\:a^3\:\Sigma^2_0\:\frac{(\frac{1}{\nu}\:\Gamma(\frac{2}{\nu}))^2}{R^*_g}
\end{equation}

\noindent
where we have used a gravitational constant $G=1$ and where the
gravitational radius $R_g$ is defined by $R_g\:=\:a\:R^*_g$. Here
$a$ is the scale parameter and $R^*_g$ is a dimensionless radius given
by the numerical approximation:
$ln(R^*_g)\:\simeq\:\frac{0.82032-0.92446\:ln(\nu)}{\nu}\:+\:0.84543$
(M\'arquez et al. 2001).

The second principle of thermodynamics states that a dynamical system
in an equilibrium state has a maximum entropy. For a self-gravitating
system in quasi-equilibrium, such as a galaxy cluster, there is no
state with an entropy maximum. On the contrary, the entropy is
constantly growing, but slowly, in a secular time scale, during which
we can consider it as nearly constant. In spite of the X-ray emission,
which is responsible in many cases for cooling flow processes which
affect the equilibrium state of the cluster, we may therefore consider
these objects as structures where dissipation processes are negligible
compared to their gravitational energy, thus settling into a
thermodynamical equilibrium.

The specific entropy of the intra-cluster gas can be obtained from the
distribution function in the phase space,
$f(\bf{\overrightarrow{x}},\bf{\overrightarrow{v}})$, of the gas
particles. The calculation is made using the microscopic
Boltzmann-Gibbs definition:

\begin{equation}
\label{boltzmann-gibbs-entropy}
s = \frac{S}{M_{gas}}= \frac{ \int f\:ln (f)\:d^3x\:d^3v }{\int f\:d^3x\:d^3v }
\end{equation}

Estimating the distribution function $f$ may be very difficult and
some hypotheses are needed to simplify the problem. The first
hypothesis is that our system has spherical symmetry, and the second
one is that the velocity dispersion of the gas particles is
isotropic. With these assumptions we are neglecting any possible
rotation of the gas, and $f$ can be obtained by Abel inversion from
the density profile (Binney \& Tremaine, 1987).  Using equations
(\ref{mm_profile}) and (\ref{rho0_sigma0}), the gas specific entropy
(Lima Neto et al., 1999; M\'arquez et al., 2000) is then:

\begin{equation}
\label{specific-entropy}
s=\frac{3}{2}\:ln\:\Sigma_0\:+\frac{9}{2}\:ln\:a\:+F(\nu)
\end{equation}

\noindent
where $F$ is a function of the $\nu$ parameter given by the numerical
approximation: 

$$F(\nu) \simeq
-0.795\:ln(\nu)-\frac{1.34}{\nu}\:+\:3.85\:\left(\frac{1}{\nu}\right)^{1.29}\:+\:ln(\Gamma[\frac{2}{\nu}])\:-\:0.822$$

\section{Results}

\subsection{Correlation of the S\'ersic parameters two by two}

\begin{figure}[h]
\begin{center}
\psfig{figure=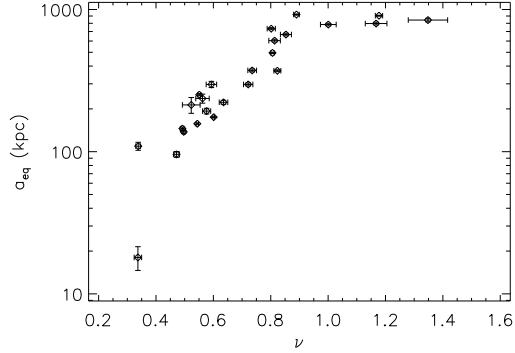,height=5truecm}
\caption{Correlation between the shape and scale parameters.
\label{fig:corr_a_nu}}
\end{center}
\end{figure}

\begin{figure}[h]
\begin{center}
\psfig{figure=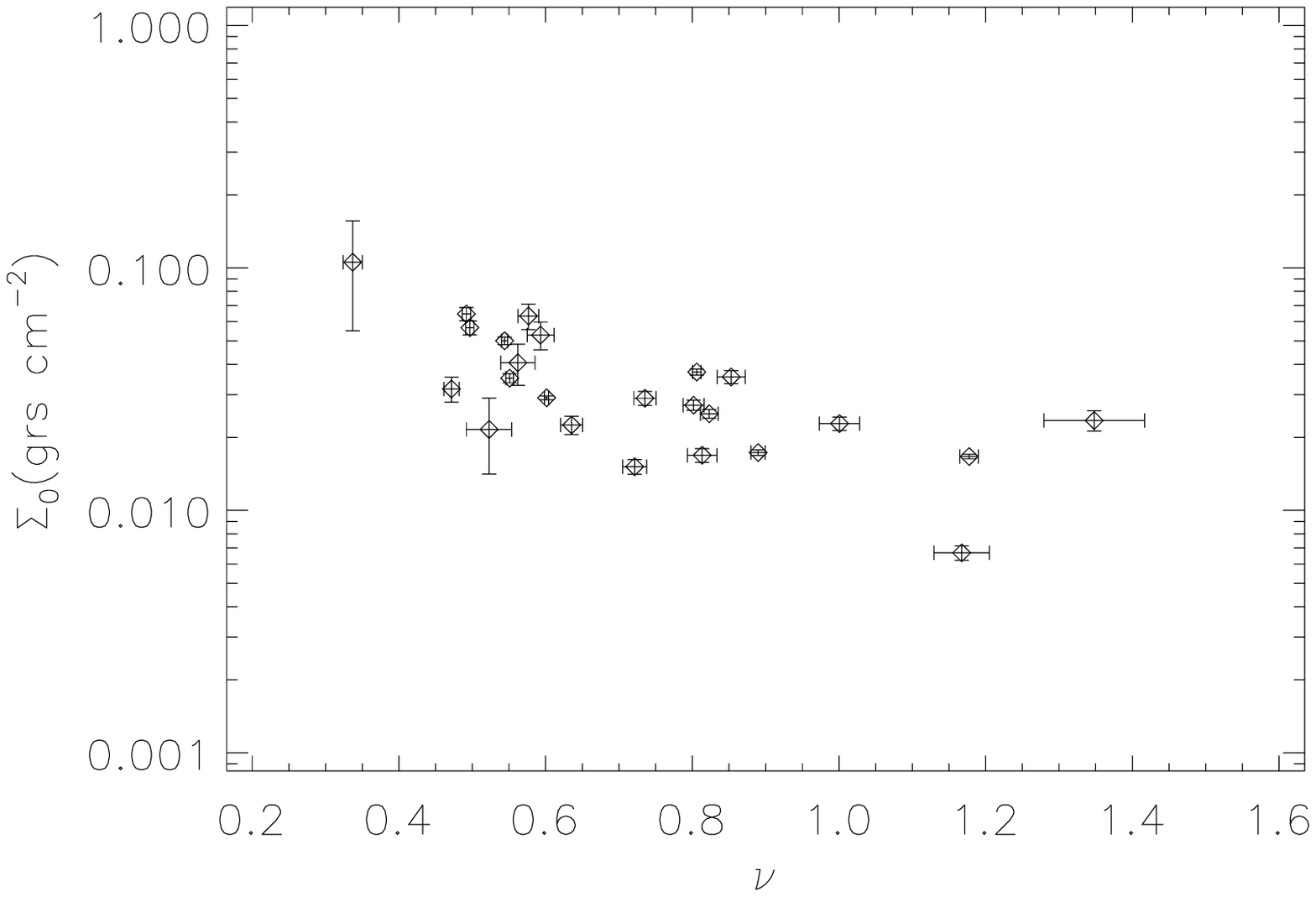,height=5truecm}
\caption{Correlation between the shape and intensity parameters.
\label{fig:corr_s0_nu}}
\end{center}
\end{figure}

\begin{figure}[h!]
\begin{center}
\psfig{figure=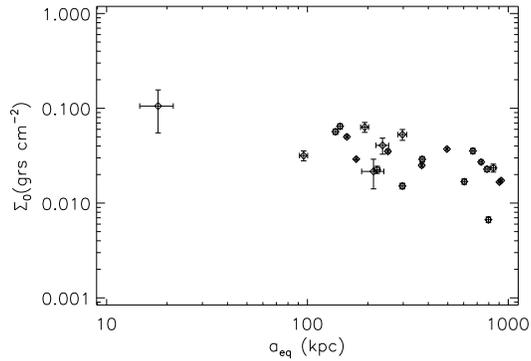,height=5truecm}
\caption{Correlation between the scale and intensity parameters.
\label{fig:corr_s0_a}}
\end{center}
\end{figure}

We see in Figs. \ref{fig:corr_a_nu}, \ref{fig:corr_s0_nu} and
\ref{fig:corr_s0_a} that the three S\'ersic parameters are correlated
two by two, as was previously observed for the optical surface
brightness of elliptical galaxies. In particular, a clear correlation
exists between the shape and scale parameters
(Fig. \ref{fig:corr_a_nu}), with a shape very similar to that found
for elliptical galaxies.  The other parameters are also correlated,
although with more dispersion than for ellipticals.

\subsection{Constancy of the specific entropy}

Once the scale parameter $a_{eq}$, the shape parameter $\nu$ and the
central density $\rho_0$ are determined, we can compute the specific
entropy $s$ for the gas component of the cluster from equation
(\ref{specific-entropy}). The gas mass can be computed from equation
(\ref{mass_gas_tot}). We can then calculate the gas entropy
$S=s\:M_{gas}$.

\begin{figure}[h]
\begin{center}
\psfig{figure=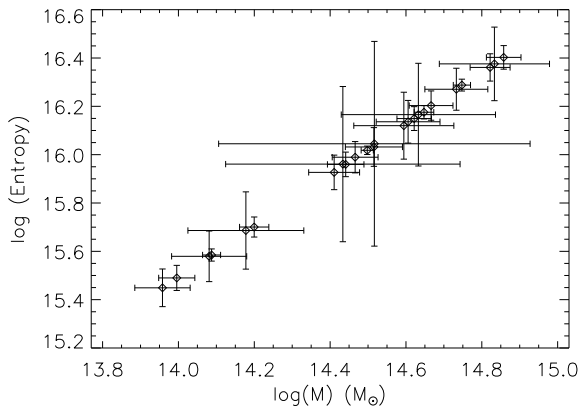,height=6truecm}
\caption{Relation between the gas entropy and mass.
\label{fig:entropy-mass}}
\end{center}
\end{figure}

A plot of the gas entropy $S$ as a function of the gas mass $M_{gas}$
is shown in Fig. \ref{fig:entropy-mass}; a linear relation exists
between the entropy and total mass of gas, implying that the specific
entropy $s$ is constant. The slope gives the value of $s\ =\ 34.5 \pm
1.3$. A remarkable characteristic of Fig. \ref{fig:entropy-mass} is
the very low dispersion of the points. 

\subsection{Scaling relation between the cluster potential energy
and the X-ray gas mass}

\begin{figure}
\begin{center}
\psfig{figure=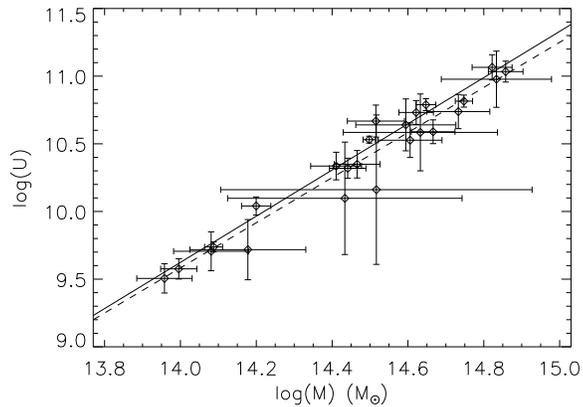,height=6truecm}
\caption{Scaling law between gas the potential energy and mass. The solid line 
corresponds to the best fit to the data and the dashed line is the 
theoretical curve (slope 5/3 in logarithmic units). 
\label{fig:energy-mass}}
\end{center}
\end{figure}

The potential energy $U_{pot}$ of the gas is displayed as a function
of the gas mass in Fig.~\ref{fig:energy-mass}.

If we write this relation as $log (U) - \alpha \,log (M) = const$, we
find a slope $\alpha = 1.72 \pm 0.05$, which is very close to the
theoretical value of $5/3$.

This result confirms the hypothesis made for the formation of
structures. The fact that elliptical galaxies and gas in clusters of
galaxies verify the scaling law between potential energy and mass
before mentioned is a strong argument in favour of the idea that
self-gravitating structures are formed by processes where the
conservation of energy and mass are verified.

\subsection{Second order relations}

Numerical simulations of elliptical galaxies formed in a hierarchical
merging scheme show that the specific entropy slightly varies with
mass, in a similar way as for observed galaxies (M\'arquez et al.,
2000). This weak dependence has been interpreted as due to merging
processes (Lima Neto et al., 1999).

\begin{figure}[h]
\begin{center}
\psfig{figure=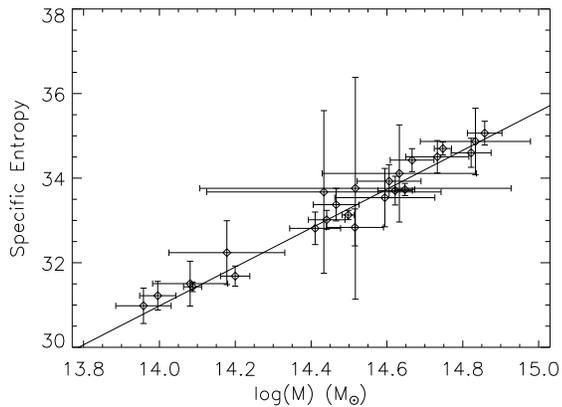,height=6truecm}
\caption{Relation between the gas specific entropy and mass.
The straight line corresponds to the best fit slope 
$\beta =4.7$.
\label{fig:s-mass}}
\end{center}
\end{figure}

We have searched for a similar relation in our sample of clusters.
The plot of the gas specific entropy $s$ versus the gas mass $M_{gas}$
shown in Fig. \ref{fig:s-mass} clearly reveals a linear relation
between $s$ and $log(M_{gas})$. The gas specific entropy $s$ is
therefore a function of the gas mass. However, this is a second order
relation compared to the more important relation $s = constant$
previously found. The difference with elliptical galaxies is that the
slope in Fig. \ref{fig:s-mass} is steeper for clusters than for
ellipticals. Thus, if we write $s=s_0+\beta\:log(M)$, we find $\beta =
4.7 \pm 0.3$ for clusters and $\beta \simeq 1$ for ellipticals.

\section{Conclusions}

The similarity of the relations found for the optical light
distribution in elliptical galaxies and for the X-ray gas emission in
clusters seems to confirm the hypothesis that the physical conditions
prevalent when these self-gravitating systems were formed are
comparable. These self-gravitating and almost dissipationless systems
are likely to have evolved in a comparable way.

The more important dependence of the specific entropy with mass for
clusters suggests that dissipating processes such as Bremsstrahlung
emission ($L \propto M^2$) and cooling flows may play an important
role as generators of entropy. Merging processes between clusters may
also be of some importance.

The present work was carried out on a sample of 24 nearby clusters;
the study of more distant objects could give us information on a
possible dependence of specific entropy with redshift and on a
possible evolution of the S\'ersic scale parameter $a$.  Such an
analysis obviously requires a higher spatial resolution and
sensitivity than those of ROSAT, and should be possible with XMM or
Chandra.



\section*{References}
\begin{thebibliography}{99}


\bibitem {} Binney J., Tremaine S. 1987, ``Galactic Dynamics'', 
Princeton University Press

\bibitem{Ebeling}Ebeling H., Voges W., B\"ohringer H. et al. 1996, 
MNRAS 281, 799

\bibitem{Gerbal}Gerbal D., Lima Neto G.B., M\'arquez I.,
Verhagen H., 1997, MNRAS 285, L41

\bibitem{James}James F. 1984, CERN Program Library Long Writeup D506

\bibitem{Jang}Jang-Condell H., Hernquist L., 2001, ApJ 548, 68

\bibitem{Lanzoni}Lanzoni B. 2000, PhD, Universit\'e Paris 7

\bibitem{Lima}Lima Neto G.B., Gerbal D., M\'arquez I., 1999, MNRAS 309, 481 

\bibitem{Marquez}M\'arquez I., Lima Neto G.B., Capelato
H., Durret F., Gerbal D., 2000, A\&A 353, 873

\bibitem{Marquezb}M\'arquez I., Lima Neto G.B., 
Capelato H., Durret F., Gerbal D., Lanzoni, B., 
2001, A\&A submitted

\bibitem{Mellier}Mellier Y., Mathez G. 1987, A\&A 175, 1

\bibitem{Sersic}S\'ersic J.L., 1968, ``Atlas de galaxias
australes, Observatorio Astron\'omico de C\'ordoba'', Argentina

\bibitem{Snowden}Snowden S. L., Kuntz K.D. US ROSAT Science Data Center, 
NASA/GSFC, 1994

\bibitem{Wu}Wu X.-P., Xue Y.-J., Fang L.-Z. 1999, ApJ 524, 22

\end{thebibliography}
\end{document}